\documentclass[12pt]{article}
\usepackage{amsmath,amssymb,amsthm,amsxtra,overpic,bm,epsfig,ulem,multirow}
\usepackage{color,cite}
\usepackage{epstopdf}
\usepackage{booktabs}
\usepackage{footnote}
\usepackage{rotating}
\usepackage{hyperref}
\usepackage{url}
\usepackage{cite}
\usepackage{tikz}
\usetikzlibrary{decorations.pathreplacing}
\usetikzlibrary{arrows,shapes,automata,backgrounds,petri}
\usetikzlibrary{decorations.pathmorphing}
\usetikzlibrary{decorations.markings}
\usepackage{lineno}
\usepackage{graphicx}
\usepackage{enumitem}

\newcommand{\Zloc}{Z_{\mathrm{local}}}
\newcommand{\Zglo}{Z_{\mathrm{global}}}

\newcommand{\Ha}{\mathrm{H}\alpha}
\newcommand{\halfchisq}{\tfrac{1}{2}\chi^{2}_{1}}

\graphicspath{{figure caption/figures/}}

\def\thefootnote{\fnsymbol{footnote}}

\newcommand{\dd}{\mathrm{d}}

\usepackage{array}

\begin{document}

\vspace{0.2cm}

\begin{center}

{\Large\bf Binary Hypothesis Testing: A Robust Framework Against the Look Elsewhere Effect}

\end{center}

\vspace{0.2cm}

\begin{center}
{\bf Han Zhang~$^{a,~b}$}~\footnote{Email: hzhang@ihep.ac.cn},
{\bf Xue-feng Ding~$^{a,}$}~\footnote{Email: dingxf@ihep.ac.cn},
{\bf Yu-Feng Li~$^{a,~b}$}~\footnote{Email: liyufeng@ihep.ac.cn},
{\bf Yi-fang Wang~$^{a}$}~\footnote{Email: yfwang@ihep.ac.cn},
{\bf Liang-jian Wen~$^{a}$}~\footnote{Email: wenlj@ihep.ac.cn},
{\bf Liang Zhan~$^{a}$}~\footnote{Email: zhanl@ihep.ac.cn},
\\
\vspace{0.2cm}
{$^a$Institute of High Energy Physics, Chinese Academy of Sciences, Beijing 100049, China}\\
{$^b$School of Physical Sciences, University of Chinese Academy of Sciences, Beijing 100049, China}
\end{center}

\vspace{1.5cm}

\begin{abstract}
In particle physics, discovery claims conventionally require an observed significance exceeding $5\sigma$. However, the interpretation of a $5\sigma$ result depends critically on the testing procedure, namely whether the hypothesis is tested at a single pre-specified point in parameter space or by scanning over a range of possible signal locations. This distinction gives rise to the look-elsewhere effect, a concept that is widely used but often interpreted as a simple penalty for scanning.
In this work, we reinterpret the look-elsewhere effect as a correction for procedural inconsistency arising when the null distribution is generated under one procedure while the test statistic is evaluated under another. Within the framework of hypothesis testing, we examine its implications and clarify the distinct roles of the look-elsewhere effect in binary and peak-search scenarios. In particular, we show that binary test is robust against the look-elsewhere effect, whereas peak searches require an explicit correction for the search over signal locations. Using a moderate trial factor of approximately 26, calibrated from the ATLAS Higgs search, we show that a $3\sigma$ global significance of peak search can correspond to approximately $4\sigma$ significance of the binary test for the same value of the observed test statistic.
This reformulation provides a clearer statistical interpretation of the look-elsewhere effect and offers a more coherent framework for understanding significance claims in particle physics.

\end{abstract}

\def\thefootnote{\arabic{footnote}}
\setcounter{footnote}{0}

\newpage

\section{Introduction}
\label{sec:introduction}

The gold standard for claiming a discovery in particle physics is a statistical significance of $5\sigma$~\cite{ParticleDataGroup:2026aaa}, corresponding to a $p$-value of approximately $3 \times 10^{-7}$.
This stringent threshold is adopted to reduce the probability of false discoveries in searches for rare or new physical phenomena.
However, the meaning of a quoted $n\sigma$ significance depends strongly on the analysis method used --- an often implicit but crucial aspect affecting how the results should be interpreted statistically and physically.
The core distinction lies in two common analysis paradigms: fixed binary hypothesis tests targeting a pre-specified parameter value, and blind peak searches that scan continuously over an extended parameter space to identify the most significant local excess.

In a fixed binary hypothesis test, the parameter point is specified a priori, so the statistical significance of an observed fluctuation can be evaluated directly.
In contrast, when scanning widely across parameters, the probability of finding a random fluctuation somewhere increases. This effect, known as the Look-Elsewhere Effect (LEE)~\cite{Gross:2010qma}, inflates the local significance and underestimates true $p$-values. Without correcting for LEE, statistical claims can be overly optimistic and lead to false discoveries~\cite{Lyons:2018gtc}.
Therefore, accounting for the LEE is essential for reliable interpretation of signals, especially in blind searches over high-dimensional data.

The LEE is widely applied in high-energy physics, most notably in the discovery of the Higgs boson~\cite{ATLAS:2012yve,CMS:2012qbp}, and its mathematical formulation through the Gross--Vitells upcrossing method is well established~\cite{Gross:2010qma, Cowan:2010js}.
In addition to the collider searches, astro-particle experiments also encounter the issue of LEE. Examples include studies on the source of high energy astrophysical neutrinos~\cite{Emig:2015dma}, and in the spectral analysis of solar neutrinos~\cite{Ranucci:2006rz}, constraining the dark matter self-annihilation cross-section through gamma-ray emissions from galaxy clusters ~\cite{Anderson:2015dpc} and searching for non-baryonic dark matter via X-ray emissions from the Milky Way ~\cite{sekiya2016search}.
A further cosmological example is the search for signatures of inflation in the primordial power spectrum~\cite{Hunt:2015iua,Fergusson:2014hya}.

However, the LEE is often misunderstood and frequently referred to as a 'penalty' or 'tax' on the parameter space scanning, suggesting that broader searches are inherently punished. While this framing is not entirely inaccurate, it overlooks a deeper insight. By comparing the scenarios of hypothesis testing for binary and peak searches, this work advocates a more refined conceptual perspective. The LEE is not an inherent feature of physical signals, but rather a statistical correction for \textit{procedural inconsistency} between the construction of null hypotheses and the evaluation of test statistics. The main conclusions are structured as the following key points:

Binary tests and blind peak searches differ fundamentally in parameter-space symmetry under null and alternative hypotheses, driving divergent statistical behaviors. Local significances are strongly procedure-dependent: a global $3\sigma$ excess in blind peak searches corresponds quantitatively to approximately $4\sigma$ in pre-specified binary hypothesis tests. The LEE correction is required exclusively for procedural mismatches between null-data generation and practical test implementation; consistent procedures eliminate the need for trial factor correction. The discovery of the Higgs boson can serve as a canonical real-world example of a procedural mismatch that has been successfully resolved through rigorous LEE calibration. In contrast, neutrino mass ordering represents the opposite extreme scenario of binary hypothesis testing, which is unaffected by LEE correction.

The remainder of this paper is organized as follows. Section~\ref{sec:formalism} introduces the statistical formalism of hypothesis testing and develops the framework of the LEE. Section~\ref{sec:numerical} provides numerical illustrations of the quantitative features of different types of hypothesis testing. Section~\ref{sec:consistency} clarifies the procedural-consistency origin of the LEE across representative test scenarios. Section~\ref{sec:higgs} applies the proposed framework to the interpretation of the Higgs boson discovery procedure. Finally, Section~\ref{sec:discussion} summarizes the key findings and provides practical guidelines for a standardized treatment of the LEE in future analyses.

\section{Formalism}
\label{sec:formalism}

\subsection{Statistical Framework}

A hypothesis test begins with two competing hypotheses. The {null hypothesis} $H_0$ typically asserts that only known (background) processes are present, while the {alternative hypothesis} $H_1$ posits the existence of a new signal in addition to the background. Given observed data $\mathbf{x}$, one constructs a {test statistic} $t(\mathbf{x})$ designed to discriminate between $H_0$ and $H_1$.

The degree of tension between the data and $H_0$ is quantified by the $p$-value, defined as the probability, under $H_0$, of obtaining a test statistic at least as extreme as the one observed:
\begin{equation}
    p = P(t \geq t_{\mathrm{obs}} \mid H_0)
    = \int_{t_{\mathrm{obs}}}^{\infty} f(t \mid H_0) \, \dd t \,,
    \label{eq:pvalue}
\end{equation}
where $f(t \mid H_0)$ is the probability density function of $t$ under $H_0$.
By convention, the $p$-value is often converted to a {significance} $Z$ defined through the inverse of the standard normal cumulative distribution:
\begin{equation}
    p = 1 - \Phi(Z) \,,
    \qquad \text{equivalently} \qquad
    Z = \Phi^{-1}(1 - p) \,,
    \label{eq:significance}
\end{equation}
where $\Phi$ is the cumulative distribution function of the standard normal
distribution. A $5\sigma$ significance thus corresponds to $p \approx
2.87 \times 10^{-7}$.

The test statistic of choice is the {profile
likelihood ratio}. Let the data be described by a probability model
$\mathcal{L}(\mathbf{x} \mid \mu, \boldsymbol{\theta})$, where $\mu$ is the {parameter of interest} (e.g., the signal strength, with $\mu = 0$
corresponding to $H_0$ and $\mu > 0$ to $H_1$) and $\boldsymbol{\theta}$
denotes {nuisance parameters} (systematic uncertainties, background
normalizations, etc.).

The profile likelihood ratio for testing $\mu = 0$ is
\begin{equation}
    q_0 = -2 \ln \lambda(0) =
    -2 \ln \frac{\mathcal{L}(\mathbf{x} \mid 0,
    \hat{\hat{\boldsymbol{\theta}}})}
    {\mathcal{L}(\mathbf{x} \mid \hat{\mu},
    \hat{\boldsymbol{\theta}})} \,,
    \label{eq:q0}
\end{equation}
where $\hat{\mu}$ and $\hat{\boldsymbol{\theta}}$ are the unconditional maximum likelihood estimators (MLEs), and $\hat{\hat{\boldsymbol{\theta}}}$ denotes the conditional MLE of $\boldsymbol{\theta}$ with $\mu$ fixed to zero. Larger values of $q_0$ indicate greater incompatibility with $H_0$.
In practice, one often imposes the constraint $\hat{\mu} \geq 0$ to avoid negative signal strength, which leads to
\begin{equation}
    \tilde{q}_0 =
    \begin{cases}
        -2 \ln \lambda(0) & \text{if } \hat{\mu} \geq 0 \,, \\
        0 & \text{if } \hat{\mu} < 0 \,.
    \end{cases}
    \label{eq:q0tilde}
\end{equation}

\subsection{Symmetric vs.\ Asymmetric Parameter Space}

We now compare the distinct properties of hypothesis testing with binary and blind peak searches.
Firstly, considering a test for the presence of a signal at a {fixed, pre-specified} point in parameter space --- for example, testing whether a particle exists at exactly $m_H = 125$~GeV~\cite{ATLAS:2012yve,CMS:2012qbp} with a known production cross-section, or determining whether the neutrino mass ordering is normal or inverted~\cite{Qian:2015waa}. 
In this case, both $H_0$ and $H_1$ are defined over the same set of parameters: the signal strength $\mu$ and the nuisance parameters $\boldsymbol{\theta}$. The only difference is that the value of $\mu$ is zero under $H_0$ and nonzero under $H_1$.
This binary hypothesis testing is featured with a \textit{symmetric parameter space} because the dimensionality of the parameter space is the same under both hypotheses.  
The test statistic $q_0$ is evaluated at the single pre-specified point, and the null distribution $f(q_0 \mid H_0)$ is obtained by generating pseudo-experiments under $H_0$ at that same point. 
In contrast, in the general case of blind peak searches, we are searching for a signal somewhere in a mass range $[m_{\min}, m_{\max}]$ without specifying the mass in advance. The alternative hypothesis $H_1$ now includes additional parameters --- the signal mass $m$, and potentially the signal width $\Gamma$ and amplitude $N$ --- that have no counterpart under $H_0$.
The test statistic is obtained by profiling the likelihood ratio over the signal-only parameters:
\begin{equation}
    q_0^{\max} = \max_{m \in [m_{\min}, m_{\max}]} \, q_0(m) \,,
    \label{eq:q0max}
\end{equation}
where $q_0(m)$ is the profile likelihood ratio evaluated at the mass $m$. This scanning test statistic identifies the most significant excess within the entire search range. Due to the profiling of signal-only parameters, there are distinct asymmetric properties between the parameter spaces of $H_1$ and $H_0$, which have profound consequences for the distribution of the test statistic under $H_0$ --- the mathematical origin of the look-elsewhere effect, which will be developed in the next section.

\subsection{The Look-Elsewhere Effect}
\label{sec:lee}

When a blind peak search is performed over a range of the signal mass parameter, the most significant excess found is, on average, larger than what one would expect from a single binary test, even in the absence of any signal. The look-elsewhere effect (LEE) quantifies this enhancement and provides a principled way to connect local and global significance.

Consider a search for a signal in the mass range $[m_{\min}, m_{\max}]$.  At each mass point $m$, one can compute the profile likelihood ratio $q_0(m)$ and the corresponding local $p$-value $p_{\mathrm{local}}(m)$. The most significant local excess observed in the scan is characterized by
\begin{equation}
    q_{0,\mathrm{obs}}^{\max}
    =
    \max_{m\in[m_{\min},m_{\max}]} q_{0}(m),
\end{equation}
or equivalently by the minimum local \(p\)-value,
\begin{equation}
    p_{\mathrm{local}}
    =
    \min_{m\in[m_{\min},m_{\max}]}
    p_{\mathrm{local}}(m).
    \label{eq:plocal}
\end{equation}
In contrast, the {global $p$-value} accounts for the fact that the search was conducted over the entire range:
\begin{equation}
    p_{\mathrm{global}} = P\!\left(
    \max_{m\in[m_{\min},m_{\max}]} q_0(m) \geq q_{0,\mathrm{obs}}^{\max}
    \;\middle|\; H_0 \right) .
    \label{eq:pglobal}
\end{equation}
Because the global $p$-value allows such an excess to occur anywhere in the scanned range, while $p_{\mathrm{local}}$ refers to the local probability at the observed point of maximum excess, we always have $p_{\mathrm{global}} \geq p_{\mathrm{local}}$.  
Equivalently, the global significance $Z_{\mathrm{global}} \leq Z_{\mathrm{local}}$. Therefore, correcting for the LEE reduces the significance. The trial factor $\tau$ is defined as the ratio of the global to local $p$-values:
\begin{equation}
    \tau = \frac{p_{\mathrm{global}}}{p_{\mathrm{local}}} \,.
    \label{eq:trial_factor}
\end{equation}
According to the Gross--Vitells upcrossing formula~\cite{Gross:2010qma}, the trial factor depends on three quantities:
\begin{itemize}
    \item The {search range} $[m_{\min}, m_{\max}]$: a wider range contains more independent fluctuations.
    \item The {mass resolution}: finer resolution increases the number of effectively independent test points.
    \item The {significance threshold}: the trial factor is itself a function of the observed significance.
\end{itemize}

In typical particle-physics searches, as well as in astrophysical and
cosmological studies, trial factors range from $\mathcal{O}(10)$ to
$\mathcal{O}(100)$ or even higher. First, the Higgs-boson searches at the
LHC involved trial factors of approximately $20$--$30$~\cite{ATLAS:2012yve,
CMS:2012qbp}. Second, in the search for gamma-ray lines toward galaxy
clusters with Fermi-LAT~\cite{Anderson:2015dpc}, the global significance is
suppressed with respect to the local one by a trial factor of
$\mathcal{O}(300)$. Third, claims of spectral features in the spectrum of
primordial density perturbations must be corrected by a LEE trial factor of
$\mathcal{O}(10^{3})$, since the reconstruction scans over a wide range of
independent wavenumbers~\cite{Hunt:2015iua,Fergusson:2014hya}.

To illustrate the local significance required to achieve a global significance of \(3\sigma\) (i.e., \(p_{\mathrm{global}} \approx 1.35 \times 10^{-3}\)) in a typical peak search, we have the following relationship:
\begin{equation}
    p_{\mathrm{local}} \approx \frac{p_{\mathrm{global}}}{\tau}
    \approx \frac{1.35 \times 10^{-3}}{20}
    \approx 6.7 \times 10^{-5} \,,
\end{equation}
which corresponds to \(Z_{\mathrm{local}} \approx 3.8\sigma\). Thus, achieving \(3\sigma\) global significance in a typical peak search requires approximately \(4\sigma\) local significance.
More generally, the relationship between \(Z_{\mathrm{local}}\) and \(Z_{\mathrm{global}}\) for representative trial factors is not a fixed offset in \(\sigma\); it depends on both the trial factor and the significance level.

\section{Numerical Illustration}
\label{sec:numerical}

In this section, quantitative features for different types of hypothesis testing will be presented by using the simulation of pseudo-experiments. 
All simulations are based on a single absorption-line spectroscopy model. The wavelength window spans $[6517.5,\,6608.5]\,\text{\AA}$, divided into $91$ equal $1\,\text{\AA}$ bins. The line resolution is set at $\sigma_{L}=5\,\text{\AA}$, with a continuum given by $c_{\mathrm{true}}=100$ counts/\text{\AA}. The nominal line center corresponds to the $\Ha$-analog wavelength $\lambda_{0}=6563\,\text{\AA}$. The search window length is selected such that the scan trial factor in the $Z_{\mathrm{th}}=3$ Gross--Vitells convention is approximately $\tau \approx 26$, which matches the ATLAS Higgs search. All pseudo-experiments with $N_{\mathrm{toys}} = 5000$ are generated from Poisson statistics without systematics.

We explore three procedurally distinct hypothesis-test configurations: 
\begin{itemize}
    \item \textbf{Case~1} (\emph{binary}), a binary case with the absorption model $\mathrm{H}_{0}: N = -N_{0}$ versus the emission model $\mathrm{H}_{1}: N = +N_{0}$ at a fixed $\lambda = \lambda_{0}$ and $N_{0} = 70$, which is tuned so that the binary Asimov separation is $A_{0} = 9$, representing the worst‑case minimum of $\mathcal{Z}_\mathrm{bin}$ at $t = 9$ given $A_{0}$;
    \item \textbf{Case~2} (\emph{global}), a peak search where the line position $\lambda$ is profiled over the full search window under $\mathrm{H}_{1}$ with $N \geq 0$;
    \item \textbf{Case~3} (\emph{local}), a single-point peak test in which $\lambda$ is fixed at $\lambda_{0}$ under $\mathrm{H}_{1}$ with $N \geq 0$.
\end{itemize}

\begin{figure}
    \centering
    \includegraphics[width=0.85\linewidth]{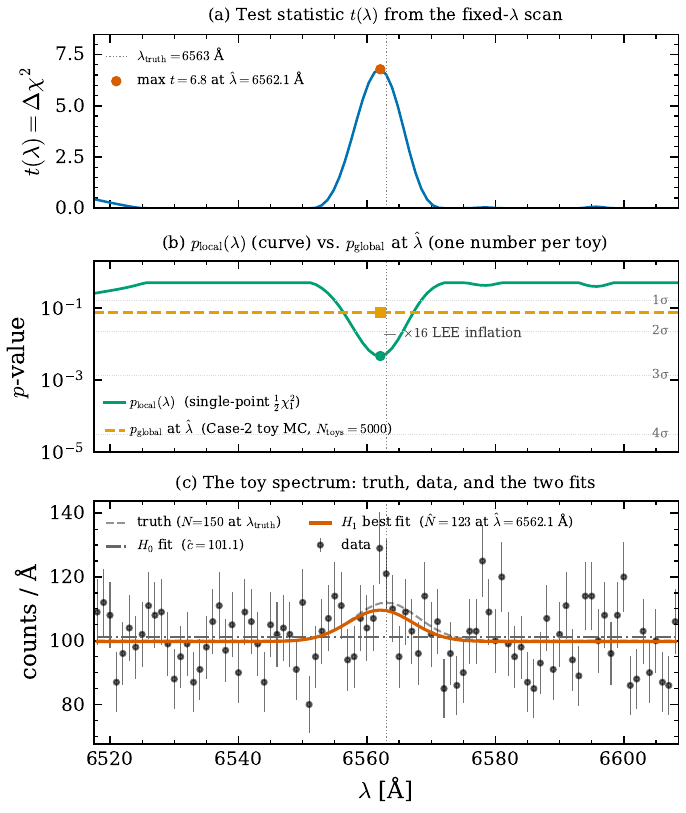}
\caption{Local and global $p$-values from a pseudo-experiment based on the $\mathrm{H}_{1}$ truth of Case~3, with a single line of amplitude $N=150$ centered at $\lambda_{0}=6563\,\text{\AA}$. Panel (a) displays the fixed-$\lambda$ test statistic $t(\lambda)=\Delta\chi^{2}(\lambda)$, peaking at $t_{\max}=6.8$ at $\hat{\lambda}=6562.1\,\text{\AA}$. Panel (b) presents the local $p$-value $p_{\mathrm{local}}(\lambda)$ (green curve) and the global $p$-value $p_{\mathrm{global}}=0.077$, indicated by a horizontal dashed orange line. The gap at $\hat{\lambda}$ represents the LEE inflation factor, which is $\times 16$. Panel (c) illustrates the spectrum, showing the noiseless truth (gray dashed), the $\mathrm{H}_{0}$ best fit (continuum only), and the $\mathrm{H}_{1}$ best fit ($\hat{N}=123$ at $\hat{\lambda}=6562.1\,\text{\AA}$).}
    \label{fig:pvalue-scan}
\end{figure}

\begin{figure}
    \centering
      \includegraphics[width=0.95\linewidth]{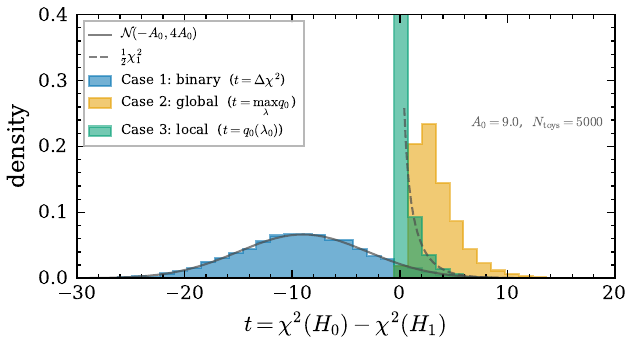}
         \caption{Empirical null distributions of the test statistic $t=\chi^{2}(\mathrm{H}_{0})-\chi^{2}(\mathrm{H}_{1})$ for three configurations of hypothesis testing. For each case, $5000$ Poisson pseudo-experiments are generated under the $\mathrm{H}_{0}$ truth, with both hypotheses fitted using profile likelihood. The gray solid curve represents the Wald-asymptotic Gaussian $\mathcal{N}(-A_{0},\,4A_{0})$ for the binary case with Asimov separation $A_{0}=9.0$, while the gray dashed curve shows the Wilks $\halfchisq$ asymptotic prediction for the single-point case.}
    \label{fig:t-distributions}
\end{figure}

In Fig.~\ref{fig:pvalue-scan} we show the operational separation of local and global $p$-values on a single pseudo-experiment from the $\mathrm{H}_{1}$ truth of \textbf{Case~3} with $N=150$ at $\lambda_{0}$.
Panel~(a) displays the fixed-$\lambda$ test statistic $t(\lambda)=\Delta\chi^{2}(\lambda)$, which is obtained by profiling the continuum $c$ at each scan position. The maximum value for this toy model is $t_{\max}=6.8$ at $\hat{\lambda}=6562.1\,\text{\AA}$. Panel~(b) illustrates the point-by-point local $p$-value $p_{\mathrm{local}}(\lambda)$, derived by applying the single-point Wilks rule $\halfchisq$ (green curve). The global $p$-value, defined as $p_{\mathrm{global}}=\Pr_{\mathrm{H}_{0}}\!\bigl(\max_{\lambda} t(\lambda)\geq t_{\max}\bigr)=0.077$, is obtained directly from the $5000$-toy empirical distribution of Case~2 depicted in Fig.~\ref{fig:t-distributions}. This value is represented as a horizontal dashed orange line, as $p_{\mathrm{global}}$ is, by design, a single number for each dataset. The vertical gap at $\hat{\lambda}$ between the green curve and the orange line indicates the LEE inflation factor, which is a multiplicative $\times 16$ in this case. Panel~(c) presents the data spectrum, including the noiseless truth, represented by the gray dashed line with $N=150$ at $\lambda_{0}$, the best fit under $\mathrm{H}_{0}$ (continuum only), and the best fit under $\mathrm{H}_{1}$ ($\hat{N}=123$ at $\hat{\lambda}=6562.1\,\text{\AA}$). This figure effectively differentiates between the local and global $p$-values based on the same dataset, visually representing the effect of LEE inflation.

\begin{figure}[!htbp]
    \centering
    \includegraphics[width=0.95\linewidth]{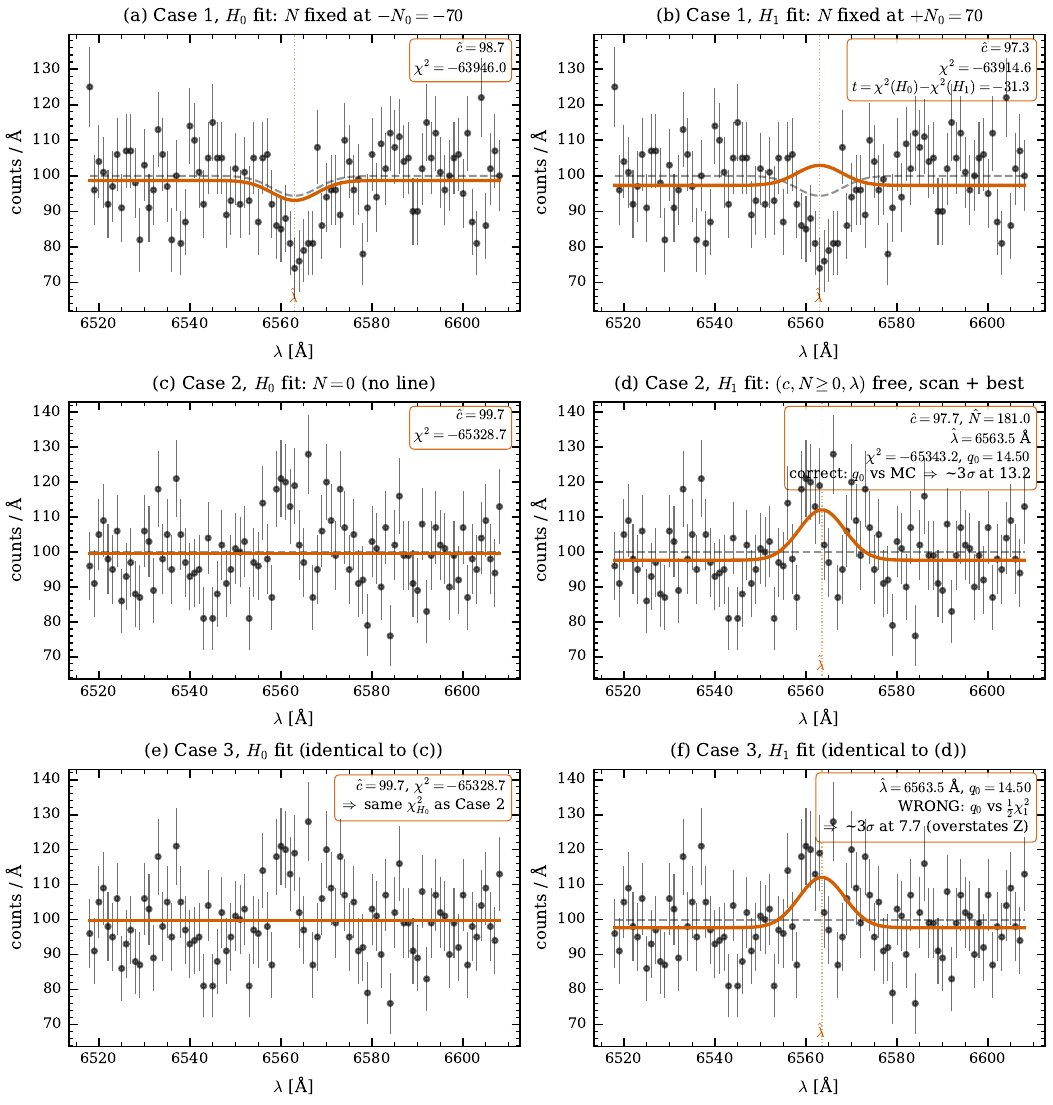}
\caption{Six example fits, one for each (case, hypothesis) combination, drawn from the toy ensembles of three hypothesis testing configurations. Each panel displays the spectrum of a pseudo-experiment (gray points with Poisson errors), the noiseless truth model (gray dashed), and the best-fit model under the indicated hypothesis (solid orange), with fit parameters and $\chi^{2}$ annotated. Rows correspond to three test configurations: Case~1 (binary test with \(N\) fixed to \(\mp N_0\)), Case~2 (global scan with \((c,N,\lambda)\) free, \(N\geq0\)), and Case~3 (local test with \(\lambda=\lambda_0\)). Columns show the corresponding \(\mathrm{H}_0\) and \(\mathrm{H}_1\) fits.}
    \label{fig:example-fits}
\end{figure}

We illustrate the example fits in Fig.~\ref{fig:example-fits} from toy ensembles of \textbf{Case~1} (upper panel), \textbf{Case~2} (middle panel), and \textbf{Case~3} (lower panel). The left and right panels display the fits for the null hypothesis ($\mathrm{H}_{0}$) and the alternative hypothesis ($\mathrm{H}_{1}$), respectively. In each panel, the spectra of the pseudo-experiments are presented as gray points with Poisson errors, along with the noiseless truth depicted by a gray dashed line, and the best-fit model for the indicated hypothesis represented by a solid orange line. The fit parameters and $\chi^{2}$ values are also provided for comparison.

Note that the $\mathrm{H}_{0}$ panels for \textbf{Case~2} and \textbf{Case~3} (panels c and e) consist of identical continuum-only fits to the same toy spectrum. The $\mathrm{H}_{1}$ panels of \textbf{Case~2} and \textbf{Case~3} (panels d and f) are also fitted to the same toy spectrum, differing only in the method of scanning $\lambda$: in panel (d), $\lambda$ is scanned over the window, while in panel (f), it is fixed at $\lambda_{0}$. Although the $\mathrm{H}_{1}$ fits in panels (d) and (f) appear visually indistinguishable, the same $\Delta\chi^{2}$ value maps to very different significances when compared with the null distributions in \textbf{Case~2} (global) versus \textbf{Case~3} (local) shown in Fig.~\ref{fig:t-distributions}. This discrepancy highlights that the LEE arises from the choice of reference distribution rather than the fitting process itself.

Fig.~\ref{fig:t-distributions} shows the empirical null distributions of the test statistic $t=\chi^{2}(\mathrm{H}_{0})-\chi^{2}(\mathrm{H}_{1})$ for the three hypothesis-test configurations, built from $5000$ Poisson pseudo-experiments per case.
The three cases share the same data and the same underlying likelihood, but differ only in which parameters are free in the fit and whether $\lambda$ is scanned or fixed.
From the figure one sees that the large-$\Delta\chi^{2}$ tail is populated most heavily by \textbf{Case~2}: for any fixed observed $\Delta\chi^{2}$, the inferred significance is therefore lowest in \textbf{Case~2} (global) and largest in \textbf{Case~1} (binary), with \textbf{Case~3} (local) in between. 
This excess weight in the right-hand tail of \textbf{Case~2} is precisely the LEE, expressed at the level of the
raw test statistic rather than the $p$-value.

\begin{figure}
    \centering
    \includegraphics[width=0.85\linewidth]{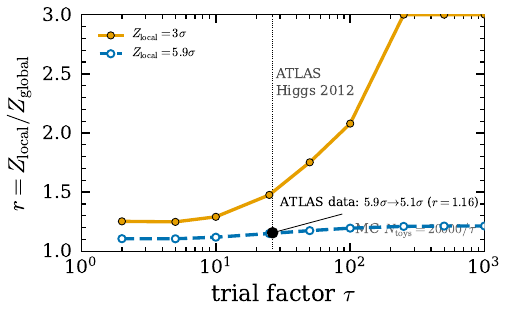}
\caption{LEE inflation factor $r=\Zloc/\Zglo$ as a function of the trial factor $\tau$. 
The points are generated using  $20,000$ pseudo-experiments of Case~2. The vertical dotted line indicates the ATLAS Higgs trial factor ($\tau \approx 26$), with the filled black dot marking the corresponding ATLAS data point on the $5.9\sigma$ curve.}
    \label{fig:r-vs-tau}
\end{figure}

The relationship between the LEE correction and the number of effectively independent search trials is illustrated in Fig.~\ref{fig:r-vs-tau}, which maps the LEE inflation factor \(r = {Z_{\mathrm{local}}}/{Z_{\mathrm{global}}}\) against the trial factor \(\tau\) at different significance levels.
For nine values of \(\tau \in \{2, 5, 10, 25, 50, 100, 250, 500, 1000\}\), the length of scan window is set to \(L = \nu_{0} \cdot 2\pi\sigma_{L}\), with \(\nu_{0}\) chosen such that the Gross–Vitells expression \(\tau = (\langle N(t_{u}) \rangle + p_{\mathrm{local}}(t_{u})) / p_{\mathrm{local}}(t_{u})\) holds for the \(Z_{\mathrm{th}} = 3\) convention~\cite{Gross:2010qma}. A new \textbf{Case~2} ensemble of \(20,000\) pseudo-experiments is generated for each \(\tau\), followed by profile fitting, to build the empirical distribution of \(\max_{\lambda} t(\lambda)\).

In the figure, the \(3\sigma\) curve (solid orange, filled circles) is extracted directly from the empirical survival function at the threshold \(t = Z_{\mathrm{local}}^{2} = 9\). The \(5.9\sigma\) curve (dashed blue, open circles), which is not reachable by Monte Carlo with the current toy count, is obtained using the Gross–Vitells upcrossing procedure, evaluated for each \(\tau\) from the same Monte Carlo configuration. We choose to plot the \(5.9\sigma\) curve rather than the conventional \(5\sigma\), in order to include the ATLAS Higgs reference point \((5.9\sigma \to 5.1\sigma, r = 1.16)\). The vertical dotted line marks the ATLAS Higgs trial factor \((\tau \approx 26)\); the filled black dot at that \(\tau\) on the \(5.9\sigma\) curve represents the ATLAS data point.
The figure quantifies the LEE as a single multiplicative correction \(r\) that depends only on \(\tau\) and the local significance threshold. It shows that inflation grows rapidly with \(\tau\) at the \(3\sigma\) threshold (from \(r \approx 1.25\) at \(\tau = 2\) up to \(r \geq 3\) at \(\tau \geq 250\)), while remaining mild at the \(5.9\sigma\) threshold \((r \approx 1.2\) across three decades of \(\tau)\).

\section{Procedural Consistency and the Origin of the LEE}
\label{sec:consistency}

In this section, we argue that the LEE arises specifically from a procedural inconsistency between the construction of the null distribution and the evaluation of the test statistic. To clarify this, we consider three scenarios for testing the existence of a signal within a wavelength range $[\lambda_{\min}, \lambda_{\max}]$. Each scenario uses the same data and likelihood model but varies the combination of null distribution generation and test statistic evaluation.

\begin{enumerate}[leftmargin=*,itemsep=0.5em]

\item\textbf{Scenario A: Consistent binary test}

In the first scenario, both the null distribution and the observed test statistic are evaluated at a single, fixed wavelength point $\lambda_0$:
\begin{enumerate}
\item \textbf{Null distribution:} Generate pseudo-experiments under $H_0$ and compute the test statistic $q_0(\lambda_0)$ for each experiment at the fixed wavelength $\lambda_0$. This results in the distribution $f(q_0(\lambda_0) \mid H_0)$.
\item \textbf{Observed test statistic:} Calculate $q_0(\lambda_0)$ from the observed data at the same wavelength $\lambda_0$.
\item \textbf{$p$-value:} Compare the observed test statistic $q_{0,\mathrm{obs}}(\lambda_0)$ to the null distribution $f(q_0(\lambda_0) \mid H_0)$.
\end{enumerate}

Because the same procedure (evaluation at $\lambda_0$) is used for both the null distribution and the observed test statistic, the $p$-value is correctly calibrated by construction, eliminating the need for a LEE correction. This constitutes a standard binary hypothesis test: the signal hypothesis is fully specified prior to examining the data, and the resulting $p$-value retains its nominal frequentist interpretation.

\item\textbf{Scenario B: Consistent peak search}

In the second scenario, both the null distribution and the observed test statistic are obtained by scanning over the full wavelength range:

\begin{enumerate}
    \item \textbf{Null distribution:} Generate pseudo-experiments under $H_0$, and for each experiment, compute \(q_0^{\max} = \max_{\lambda} q_0(\lambda)\) by scanning over the range \([\lambda_{\min}, \lambda_{\max}]\). This results in the distribution \(f(q_0^{\max} \mid H_0)\).
    \item \textbf{Observed test statistic:} Compute \(q_{0,\mathrm{obs}}^{\max} = \max_{\lambda} q_0(\lambda)\) from the observed data by scanning over the same wavelength range.
    \item \textbf{$p$-value:} Compare \(q_{0,\mathrm{obs}}^{\max}\) to \(f(q_0^{\max} \mid H_0)\).
\end{enumerate}

The same scanning procedure is used for both the null distribution and observed test statistic. The null distribution of \(q_0^{\max}\) is generally shifted towards larger values and exhibits a heavier right-hand tail compared to the single-point distribution \(f(q_0(\lambda_0) \mid H_0)\). This effect arises directly from maximizing the local test statistic over the scanned wavelength range. However, this is already accounted for in \(f(q_0^{\max} \mid H_0)\). When comparing the observed value \(q_{0,\mathrm{obs}}^{\max}\) with this null distribution, the resulting \(p\)-value is a properly calibrated global \(p\)-value. No additional trial factor correction is required because the scanning procedure has been incorporated into the construction of the null distribution.

\item\textbf{Scenario C: Inconsistent mixed procedure}

The third scenario, often encountered as an intermediate step in practical searches, combines the two procedures:

\begin{enumerate}
    \item \textbf{Null distribution:} Use the single-point distribution \(f(q_0(\lambda_0) \mid H_0)\) from Scenario~A, or equivalently, the analytic \(\chi^2\) approximation derived from Wilks' theorem.
    \item \textbf{Observed test statistic:} Compute \(q_{0,\mathrm{obs}}^{\max} = \max_{\lambda} q_0(\lambda)\) by scanning over the range \([\lambda_{\min}, \lambda_{\max}]\), following the procedure from Scenario~B.
    \item \textbf{$p$-value:} Compare the observed scanning test statistic to the single-point null distribution.
\end{enumerate}

Mis-calibration occurs when the scan-selected statistic \(q_{0,\mathrm{obs}}^{\max}\) is compared to the single-point null distribution \(f(q_0(\lambda_0) \mid H_0)\). The probability of finding a significant fluctuation somewhere within the scanned range is omitted, leading to a \(p\)-value that is too small and an overestimation of significance.
The LEE correction restores the corresponding global \(p\)-value, achieved either through a trial factor, the Gross Vitells upcrossing approximation, or by directly constructing the null distribution of \(q_0^{\max}\).

\end{enumerate}

\begin{figure}
    \centering
    \includegraphics[width=0.9\linewidth]{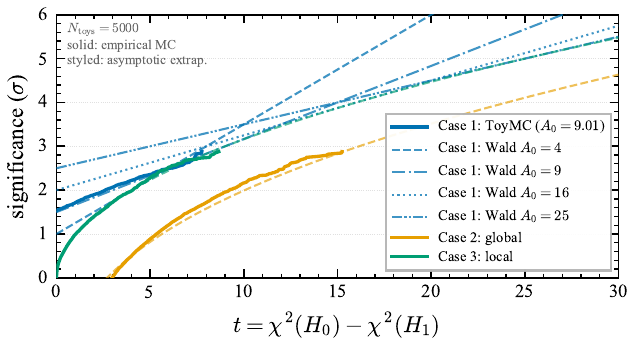}
    \caption{Empirical mapping (solid) from the test statistic \( t \) to the one-sided significance \( Z \) for all three cases, with corresponding asymptotic extrapolations (styled) extending into the deep tail beyond the brute Monte Carlo reach. The empirical survival \( S(t) \) is converted to \( Z(t) \) and plotted as solid lines in respective colors: blue for Case~1 (binary), orange for Case~2 (global), and green for Case~3 (local). Styled lines represent the appropriate asymptotic extrapolations. }
    \label{fig:t-to-z}
\end{figure}

The three scenarios are illustrated in Fig.~\ref{fig:example-fits}, featuring six example fits drawn from the toy ensembles shown in Fig.~\ref{fig:t-distributions}, with one fit corresponding to each (case, hypothesis) pair. The key point is that the visually indistinguishable \(\mathrm{H}_{1}\) fits in the \textbf{Case~2} and \textbf{Case~3} rows yield significantly different significances when evaluated against their respective null distributions. This discrepancy highlights that the LEE arises from the choice of reference distribution rather than from the fit itself.

Fig.~\ref{fig:t-to-z} illustrates the empirical mapping from the test statistic \( t=\chi^2(H_{0})-\chi^2(H_{1}) \) (denoted as $\Delta\chi^2$) to the one-sided significance \( Z \) for the three cases, overlaid with the case-appropriate asymptotic extrapolation that extends each curve into the deep tail beyond the reach of brute Monte Carlo methods.

For each of three null distributions shown in Fig.~\ref{fig:t-distributions}, the empirical survival function \( S(t) = \#\{t_{k} \geq t\} / N_{\mathrm{toys}} \) is converted to \( Z(t) = \Phi^{-1}(1 - S(t)) \) and plotted as a solid line in the corresponding case color (blue for \textbf{Case~1} (binary), orange for \textbf{Case~2} (global), green for \textbf{Case~3} (local)). The curve is truncated where fewer than 10 toys remain above \( t \), i.e., with \( N_{\mathrm{toys}} = 5000 \), this limits the empirical resolution to \( Z \approx 3\sigma \).
The styled continuation in the same color represents the appropriate asymptotic extrapolation.
The case-by-case asymptotic formulas are as follows: for \textbf{Case~1}, the Wald–Gaussian \( Z_{\mathrm{binary}}(t) = (t + A_{0}) / \sqrt{4A_{0}} \) evaluated at both edges of the curves, where $A_{0}$ is the Asimov separation; for \textbf{Case~2}, the Gross–Vitells upcrossing formula with \( \nu_{0}^{\mathrm{MC}} = 2.62 \) calibrated on the \textbf{Case~2} toy array; and for \textbf{Case~3}, the one-sided Wilks \( Z_{\mathrm{local}}(t) = \sqrt{t} \) derived from \( \frac{1}{2} \chi^{2}_{1} \).
In \textbf{Case~2} (global) and \textbf{Case~3} (local), the empirical and dashed curves overlap, demonstrating that the asymptotic formulas are correctly calibrated for this toy. In \textbf{Case~1} (binary), the empirical and asymptotic curves also overlap when the observed \( t \) is near the Asimov \( A_{0} \). It is important to note that regardless of the chosen \( A_{0} \) value, the significance of \textbf{Case~1} (binary) is always higher than that of \textbf{Case~3} (local).

By comparing the \textbf{Case~1} (binary) and \textbf{Case~2} (global) curves at the same observed \( \Delta\chi^{2} \), it is evident that the binary test consistently yields a significantly higher \( Z \) than the peak-search test. The gap between the blue and orange curves serves as an operational statement of how much additional significance a true binary configuration provides over a wide-band scan on the same data. It should be emphasized that this additional significance is dependent on the value of the trial factor \( \tau \), with a moderate \( \tau = 26 \) being calibrated from the ATLAS Higgs search.

Although accurate \(p\)-values are produced for Scenario~B without requiring any LEE correction, experiments typically favor Scenario~C due to its lower computational cost. In Scenario~B, the null distribution \(f(q_0^{\max} \mid H_0)\) must be constructed by scanning over the full mass range for each pseudo-experiment.
For a search involving \(N_m\) parameter points and \(N_{\mathrm{PE}}\) pseudo-experiments, this necessitates \(N_m \times N_{\mathrm{PE}}\) likelihood evaluations, each of which may involve a computationally intensive fit. In the case of the Higgs search, where \(N_m \sim \mathcal{O}(100)\) and the requirement is \(N_{\mathrm{PE}} \gg 10^7\) to probe the \(5\sigma\) tail, this approach becomes prohibitively expensive.
Scenario~C, in contrast, is significantly more efficient. The single-point null distribution can be computed either analytically via Wilks' theorem or from a modest number of pseudo-experiments at a single mass point. The scanning test statistic is computed only once on the observed data.
The LEE correction bridges this gap at negligible additional cost, in which the Gross–Vitells method requires only \(\langle N_u(0) \rangle\), which can be estimated from \(\mathcal{O}(100)\) pseudo-experiments at a low threshold. 

Therefore, the essential insight of this section is that the LEE is not merely a cost associated with scanning; rather, it serves as a correction for procedural inconsistency. It arises because experiments, for practical reasons, generate null distributions using one procedure with a single parameter point while evaluating the test statistic using another scanning method. The LEE correction reconciles these two different procedures, recovering the outcome that a fully consistent approach would have yielded.

\section{Application to the Higgs Discovery}
\label{sec:higgs}

The discovery of the Higgs boson in 2012 by the ATLAS~\cite{ATLAS:2012yve} and CMS~\cite{CMS:2012qbp} experiments at the Large Hadron Collider serves as a canonical illustration of the LEE and the procedural consistency framework developed in Sec.~\ref{sec:consistency}.

Both ATLAS and CMS conducted searches for the Higgs boson by scanning over a broad mass range. The ATLAS search covered the mass range \( m_H \in [110, 600] \) GeV, while CMS explored a similar range.

At each hypothesized mass \( m_H \), the experiments performed profile likelihood analyses that combined multiple production and decay channels, including \( H \to \gamma\gamma \), \( H \to ZZ^* \to 4\ell \), and \( H \to WW^* \to \ell\nu\ell\nu \), among others. This approach enabled the computation of the local test statistic \( q_0(m_H) \). The most significant excess in both experiments was observed at approximately \( m_H \approx 125 \) GeV, primarily driven by the \( H \to \gamma\gamma \) and \( H \to ZZ^* \to 4\ell \) channels, which provide the best mass resolution.

The reported local significances at the best-fit mass were:
\begin{itemize}
    \item ATLAS: \( Z_{\mathrm{local}} = 5.9\sigma \)~\cite{ATLAS:2012yve},
    \item CMS: \( Z_{\mathrm{local}} = 5.0\sigma \)~\cite{CMS:2012qbp}.
\end{itemize}
After accounting for the LEE over the full search mass range, the global significances were:
\begin{itemize}
    \item ATLAS: \( Z_{\mathrm{global}} = 5.1\sigma \)~\cite{ATLAS:2012yve},
    \item CMS: \( Z_{\mathrm{global}} = 4.6\sigma \)~\cite{CMS:2012qbp}.
\end{itemize}
It is important to note that the LEE correction is modest, despite the extensive search range, because the significance is already very high. The trial factor influences the \( p \)-value multiplicatively, but at \( 5\sigma \), the \( p \)-value is extremely small; thus, even a LEE inflation factor of 20–30 does not dramatically alter \( Z \).

Using the framework of Sec.~\ref{sec:consistency}, the Higgs discovery corresponds to Scenario~C:
\begin{enumerate}
    \item \textbf{Null distribution:} At each mass point $m_H$, the experiments used the asymptotic $\chi^2$ approximation (or dedicated single-point pseudo-experiments) to characterize the null distribution $f(q_0(m_H) \mid H_0)$. {This is the binary-test null}.
    \item \textbf{Test statistic:} The discovery claim was based on the most significant excess found anywhere in the search range --- effectively $q_0^{\max}$. This is the scanning test statistic.
    \item \textbf{Mismatch and correction:} The local significance
    ($5.9\sigma$ or $5.0\sigma$) was computed by comparing the scanning
    result to the single-point null. The LEE correction (Gross--Vitells
    method) then converted this to a global significance, recovering what a
    fully consistent scanning null would have produced.
\end{enumerate}

The trial factors for the Higgs search were approximately 20 to 30, depending on the search range, the mass resolution of the combined channels, and the significance threshold used in the Gross–Vitells formula. This indicates that the global \( p \)-value was roughly 20 to 30 times larger than the local \( p \)-value.
To provide context, a trial factor of 26 suggests that the broad search effectively included about 26 independent binary tests. At the mass resolution of the \( H \to \gamma\gamma \) channel (approximately 1 to 2 GeV), the search range of about 500 GeV comprises many more mass points. However, adjacent tests are highly correlated, resulting in an effective number of independent trials that is much smaller than the total number of mass points.
The relatively modest size of the trial factor for the Higgs search reflects this balance: the broad search range favors a larger trial factor, while the high mass resolution leads to strong correlations among fluctuations across nearby mass points, thereby reducing the effective number of the independent trials.

\begin{figure}
    \centering
        \includegraphics[width=0.85\linewidth]{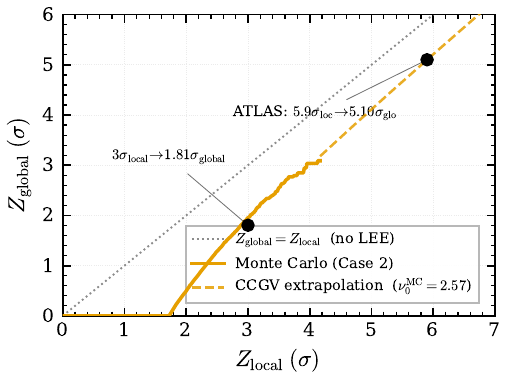}
    \caption{%
        Global versus local significance for a peak-search scan with the trial factor of \( \tau \approx 26 \). The parametric curve \( (Z_{\mathrm{local}}(t), Z_{\mathrm{global}}(t)) \) illustrates the relationship between local and global significances. The solid orange line represents the direct Monte Carlo readout, while the dashed orange line shows the LEE-corrected global significance from the Gross–Vitells extrapolation. The dotted diagonal \( y=x \) marks the LEE-free baseline.}
    \label{fig:zlocal-zglobal}
\end{figure}

Fig.~\ref{fig:zlocal-zglobal} presents the LEE as a single parametric curve displaying \( Z_{\mathrm{global}} \) versus \( Z_{\mathrm{local}} \) for a peak search using the toy's trial factor \( \tau \approx 26 \).
The parametric curve \( \bigl(Z_{\mathrm{local}}(t), Z_{\mathrm{global}}(t)\bigr) \) is generated by setting the naive single-point significance to the one-sided Wilks value \( Z_{\mathrm{local}}(t) = \sqrt{t} \) (Case~3, \(\frac{1}{2} \chi^{2}_{1}\) asymptotic) for each \( t \) and reading the LEE-corrected \( Z_{\mathrm{global}} \) from the empirical Case~2 toy mapping in Fig.~\ref{fig:t-to-z} (orange curve).
The solid orange portion represents the direct toy Monte Carlo readout (valid up to \( Z_{\mathrm{local}} \approx 3\sigma \), where at least five toys remain above the threshold in \( t_{2} \)); the dashed orange continuation is the Gross–Vitells extrapolation \( p_{\mathrm{global}}(t) = p_{\mathrm{local}}(t) + \langle N(t) \rangle \) with \( \nu_{0}^{\mathrm{MC}} = 2.57\), matching the ATLAS conversion. The dotted diagonal \( y = x \) marks the LEE-free baseline, where the multiplicative correction \( r = Z_{\mathrm{local}} / Z_{\mathrm{global}} \) would be unity. Two filled black markers anchor the chart: at \( Z_{\mathrm{local}} = 3\sigma \), the LEE reduces the global significance to approximately \( 1.8\sigma\), while the ATLAS Higgs reference point \( 5.9\sigma_{\mathrm{local}} \to 5.1\sigma_{\mathrm{global}} \) sits exactly on the extrapolated tail, according to the design of \( \nu_{0}^{\mathrm{MC}} \). The vertical gap between the curve and the diagonal represents the LEE correction \( Z_{\mathrm{local}} - Z_{\mathrm{global}} \) at that local significance, directly indicating how much an over-quoted local significance diminishes once the trial factor is accounted for.

\section{Conclusion}
\label{sec:discussion}

In this work, we have examined the look-elsewhere effect (LEE) and its implications within the framework of hypothesis testing in particle physics. The LEE is reinterpreted as a correction for procedural inconsistencies that arise when generating null distributions and evaluating test statistics.
We clarify the distinct role of the LEE in the binary test and peak-search scenarios. In particular, we show that binary test is robust against the LEE, whereas peak searches require an explicit correction for the search over signal locations.
Through a detailed analysis, we have established the following conclusions:

\begin{itemize}
    \item \textbf{Parameter-space asymmetry.}  
    The fundamental difference between a binary hypothesis test and a peak search lies in the symmetry of their parameter spaces. In a binary test, both \( H_0 \) and \( H_1 \) are defined over the same set of parameters. In contrast, a peak search introduces signal-only parameters in \( H_1 \) that are absent under \( H_0 \). This asymmetry results in the breakdown of Wilks' theorem for the scanning test statistic, which is the mathematical basis for the LEE.

    \item \textbf{Procedural inconsistency.}  
In peak search scenarios, the LEE is not an inherent penalty for broad signal searches. Instead, it arises when the null distribution is generated using one method (e.g., binary or single-point testing) while the test statistic is evaluated using a different method (e.g., scanning over a parameter range). When these two methods are aligned, as in Scenario~B, no correction for the LEE is necessary.
    
    \item \textbf{The $3\sigma \approx 4\sigma$ correspondence.} 
    A peak search with a typical trial factor of approximately 20 requires roughly \(4\sigma\) local significance to achieve \(3\sigma\) global significance. This illustrates the quantitative impact of the LEE in the context most relevant to evidence and discovery claims. In contrast, as shown in Fig.~\ref{fig:t-to-z}, the binary hypothesis test is immune to the LEE. Therefore, it can achieve a higher significance than the peak search for the same value of the test statistic \(t\).

    \item \textbf{The Higgs discovery.} 
    The ATLAS and CMS Higgs analyses serve as illustrative examples of Scenario~C, in which the experiments generated per-mass-point null distributions while evaluating a scanning test statistic. The LEE correction, applied using the Gross–Vitells formula, reconciled this mismatch, reducing the significance from local values of \(5.9\sigma\) and \(5.0\sigma\) to global values of \(5.1\sigma\) and \(4.6\sigma\).
\end{itemize}

Based on the framework discussed in this work, we present the following practical recommendations for experimentalists and analysts.

\begin{itemize}
\item Firstly, whenever possible, generate the null distribution using the same procedure employed to derive the test statistic. If the test statistic involves scanning, the null distribution should likewise be constructed from scanned pseudo-experiments. This approach eliminates the need for any LEE correction (Scenario~B).

\item Secondly, when computational constraints prevent a fully consistent scanning null distribution—particularly at high significance thresholds—employ the Gross–Vitells upcrossing formula to reconcile the differences between the single-point null and the scanning test statistic. This requires estimating \( \langle N_u(0) \rangle \) from a modest number of pseudo-experiments conducted at a lower threshold.

\item Thirdly, ensuring transparency in reporting both the local and global significances, allows readers to assess the impact of the LEE and to evaluate the evidence within the appropriate context. A local significance value alone is insufficient for a scanning search, while reporting only global significance may obscure the strength of the signal at the best-fit point.

\item Finally, including the trial factor \( \tau \) alongside local and global significances provides a concise summary of the magnitude of the LEE and facilitates comparisons across different searches with varying ranges and resolutions.
\end{itemize}

The one-dimensional scanning framework discussed in this work naturally extends in several directions. When scanning over multiple signal parameters simultaneously (e.g., mass and width, or mass and cross-section)~\cite{Vitells:2011ca}, the upcrossing formula generalizes to the expected Euler characteristic of excursion sets, as developed in Ref.~\cite{Gross:2010qma}. The trial factor increases with the dimensionality of the search space and can be related to the ratio of prior volumes under \(H_0\) and \(H_1\), thus connecting the frequentist LEE to Bayesian model comparison. A broad search range corresponds to a diffuse prior on the signal parameters, which "penalizes" \(H_1\) in the Bayes factor—an effect analogous to the frequentist trial factor~\cite{Trotta:2008qt,Bayer:2020pva}. The LEE exemplifies the broader statistical problem of multiple testing. Other corrections, such as the Bonferroni method~\cite{Dunn1961}, false discovery rate control~\cite{Benjamini:1995ram}, and closed testing procedures~\cite{MarcusPeritzGabriel1976}, address related but distinct issues. The Gross–Vitells approach is specifically tailored to the structure of particle physics searches, where the test statistics are correlated across the parameter space.

To conclude, the LEE plays a central role in discovery claims within particle physics. However, its conceptual foundations are sometimes obscured by the technical details of its implementation. By reframing the LEE as a correction for procedural inconsistency rather than as an inherent penalty for scanning, we clarify both when the correction is necessary—specifically, when null and test procedures differ—and why it is effective, as it maps the mismatched result to the correct, fully consistent one. This perspective on procedural consistency may also prove valuable in other fields where scanning or multiple testing is prevalent, such as genomics~\cite{Dudoit:2003} and gravitational-wave astronomy~\cite{LIGOScientific:2016vbw}.

\section*{Acknowledgements}
This work is supported in part by the National Key Research and Development Program of China under Grant Nos.~2024YFE0110501,~2022YFA1602000, by the
National Natural Science Foundation of China under Grant Nos.~ 12125506, 12075255.

\bibliographystyle{unsrt}
\bibliography{references}

\end{document}